\documentclass[conference]{IEEEtran}
\IEEEoverridecommandlockouts

\usepackage[utf8]{inputenc}
\usepackage{color}
\usepackage{amsmath,amssymb,mathtools,dsfont,bm,amsthm}
\usepackage{algorithm,algorithmic}
\usepackage{cite}
\usepackage{siunitx}

\usepackage[skip=-3pt]{subcaption}
\usepackage{subfloat}
\IEEEsettopmargin{t}{54pt}

\makeatletter
\renewenvironment{thebibliography}[1]{%
	\@xp\section\@xp*\@xp{\refname}%
	\normalfont\footnotesize\labelsep .5em\relax
	\renewcommand\theenumiv{\arabic{enumiv}}\let\p@enumiv\@empty
	\vspace*{-2pt}
	\list{\@biblabel{\theenumiv}}{\settowidth\labelwidth{\@biblabel{#1}}%
		\leftmargin\labelwidth \advance\leftmargin\labelsep
		\usecounter{enumiv}}%
	\sloppy \clubpenalty\@M \widowpenalty\clubpenalty
	\sfcode`\.=\@m
}{%
	\def\@noitemerr{\@latex@warning{Empty `thebibliography' environment}}%
	\endlist
	\setlength{\parskip}{-.5pt}
	\setlength{\itemsep}{-1pt}
}
\makeatother
\let\OLDthebibliography\thebibliography
\renewcommand\thebibliography[1]{
	\OLDthebibliography{#1}
	\setlength{\parskip}{-.2pt}
	\setlength{\itemsep}{-.6pt}
}

\usepackage{thmtools}

\declaretheoremstyle[%
  spaceabove=-6pt,%
  spacebelow=6pt,%
  headfont=\normalfont\itshape,%
  postheadspace=1em,%
  qed=\qedsymbol%
]{mystyle}

\title{Optimizing the Age of Information in Mixed-Critical Wireless Communication Networks}
\author{\IEEEauthorblockN{Robert-Jeron Reifert, Stefan Roth, and Aydin Sezgin}
\IEEEauthorblockA{Institute of Digital Communication Systems, Ruhr University Bochum, Bochum, Germany\\
Email: \{robert-.reifert,stefan.roth-k21,aydin.sezgin\}@rub.de}
\thanks{This work has been submitted to the IEEE for possible publication. Copyright may be transferred without notice, after which this version may no longer be accessible.\newline
This work was funded in part by the German Federal Ministry of Education and Research (BMBF) in the course of the 6GEM Research Hub under Grant 16KISK037, and in part by the Deutsche Forschungsgemeinschaft (DFG, German Research Foundation) under Germany's Excellence Strategy - EXC 2092 CASA - 390781972}
}
\date{December 2021}
\begin{document}
\bstctlcite{IEEEexample:BSTcontrol}

\maketitle

\begin{abstract}
	Beyond fifth generation wireless communication networks (B5G) are applied in many use-cases, such as industrial control systems, smart public transport, and power grids. Those applications require innovative techniques for timely transmission and increased wireless network capacities. Hence, this paper proposes optimizing the data freshness measured by the age of information (AoI) in dense internet of things (IoT) sensor-actuator networks. Given different priorities of data-streams, i.e., different sensitivities to outdated information,
	mixed-criticality is introduced by analyzing different functions of the age, i.e., we consider linear and exponential aging functions. An intricate non-convex optimization problem managing the physical transmission time and packet outage probability is derived. Such problem is tackled using stochastic reformulations, successive convex approximations, and fractional programming, resulting in an efficient iterative algorithm for AoI optimization. Simulation results validate the proposed scheme's performance in terms of AoI, mixed-criticality, and scalability.
	The proposed non-orthogonal transmission is shown to outperform an orthogonal access scheme in various deployment cases. Results emphasize the potential gains for dense B5G empowered IoT networks in minimizing the AoI.
\end{abstract}
	
\section{Introduction}
\begin{figure}
	\centering
	\includegraphics[width=0.8\linewidth]{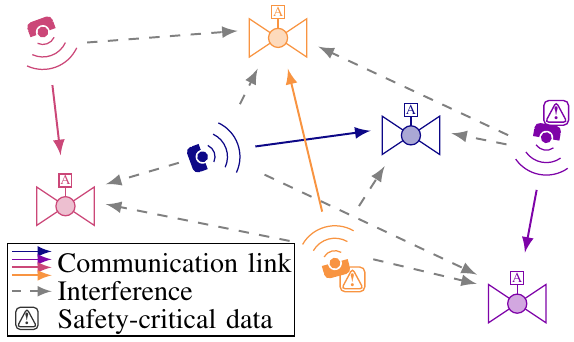}
	\vspace*{-.25cm}
	\caption{Network of sensor-actuator pairs.}
	\label{fig:network}
\end{figure}

\vspace{-.2cm}
While todays internet of things (IoT) technologies, such as massive and cellular IoT, experienced enormous growth of connected devices during the recent years, researchers forecast the total IoT connections to rise from $14.6$ billion in $2022$ to $30.2$ billion in $2027$ \cite{emr}. In wireless IoT sensor networks, usually big numbers of low-complexity sensors gather data and transmit them to various destinations, e.g., corresponding actuators.
For instance, the process industry, where uninterrupted and safe operations are vital, relies on sensors gathering data like pressure, temperature, cleanliness, etc. As such data is transmitted to different instances, e.g., the process control system or the maintenance personnel, naturally, different data-streams are of different criticality \cite{10.1145/3131347,9813735}. These IoT networks are referred to as mixed-critical regarding the communication. As an example, consider the smart public transport use-case, where sensors gather location, direction and status \cite{emr}. Data regarding human safety is considered to be safety-critical, while fuel consumption might take a subordinated role. An illustration of such mixed-critical IoT network is provided in Fig.~\ref{fig:network}.
In such safety- and time-critical applications, it is especially relevant to keep the data at the receiver fresh. As a measure for the data freshness, the \emph{age of information (AoI)} has been widely used as a performance metric \cite{9380899}.\\
\indent In this work, both the communication time and packet outage as parts of the AoI metric in mixed-critical IoT device to device (D2D) sensor-actuator networks are optimized in order to enhance the data freshness. Recent works on the AoI metric related to this paper include \cite{9380899,9174163,9302619,9646490}. The work \cite{9380899} provides an overview on the AoI metric from a general point of view. In \cite{9174163}, the authors minimize the AoI using a hybrid time-division/non-orthogonal multiple access (TDMA/NOMA) scheme. More specifically, using a Markov decision process, optimal policies for switching between TDMA and NOMA to minimize the weighted sum-AoI in a two user network are derived. While, including mixed-criticality through adapting weights is possible, such aspect is not considered in \cite{9174163}. In contrast, \cite{9302619} conducts resource allocation in heterogeneous IoT under minimizing linear and non-linear AoI. Mixed-criticality is represented in different aging functions, i.e., linear and exponential. The work \cite{9302619}, however, considers orthogonal resource blocks and scheduling decisions, whereas we herein consider a physical layer perspective under NOMA given the scheduled devices. In the same vein, the authors in \cite{9646490} optimize both data freshness and monitoring accuracy using TDMA scheduling.\\
\indent Contrary to the related works, we herein consider an IoT network of mixed-critical sensor-actuator pairs which transmit simultaneously using a D2D NOMA scheme. On the physical layer, a central controller is able to adjust data rates in order to adjust the transmission latency and outage probability. With packet failures occurring with a certain probability, we derive a closed-form analytical expression of different functions of the peak AoI representing different levels of criticality. 
Subsequently, an intricate non-convex optimization problem is formulated, which minimizes the users' sum mixed-critical peak linear and exponential AoI by jointly optimizing transmission time and outage probability. Using successive convex approximation (SCA) and  fractional programming (FP) techniques, a low-complexity iterative algorithm  for optimizing the AoI is derived. Comparing the proposed non-orthogonal to an orthogonal access scheme, numerical simulations validate the enhanced performance in terms of AoI and scalability.
\vspace{-.1cm}
\section{System Model}\label{sec:sysmod}\vspace{-.1cm}
    This paper considers a D2D network consisting of $K$ sensor-actuator pairs, as depicted in Fig.~\ref{fig:network}. In this context, this paper assumes mixed-critical sensor-actuator pairs, i.e., the data of sensor $k$ is either safety-critical or non-critical. Safety-critical D2D pairs require timely and reliable transmission, as compared to non-critical pairs. Hence, safety-critical pairs are more sensitive to outdated information, which is captured in this paper's objective in Section~\ref{sec:prob}.
    Each device, i.e., sensor or actuator, is assumed to be equipped with a single antenna, and the transmit bandwidth is $B$. Following the NOMA scheme, the $K$ sensors send their signals simultaneously via the wireless medium. Hereby, the channel from any sensor $i$ to actuator $k$ is denoted by $h_{i,k}$. The signal transmitted by sensor $k$ to its corresponding actor is indicated as $s_k$, where $s_k\sim\mathcal{CN}(0,1)$ is independent identically distributed and circularly symmetric. $\sqrt{q_k}$ refers to the corresponding transmit power. Hence, the received signal at actuator $k$ can be mathematically defined as\vspace*{-.25cm}
    \begin{align}\label{eq:yk}
	    y_k = h_{k,k} \sqrt{q_k} s_k + \sideset{}{_{i\neq k}}\sum h_{i,k} \sqrt{q_i} s_i + n_k,
	\end{align}
	\\\phantom{a}\vspace*{-.975cm}\\
	where $n_k\sim\mathcal{C}\mathcal{N}(0,\sigma^2)$ is additive white Gaussian noise with variance $\sigma^2$. Specifically, we assume $h_{i,k}$, i.e., the channel coefficients to have the form $h_{i,k} = d_{i,k}^{-\mu/2} g_k$, where $d_{i,k}$ is the distance of the sensor $i$ and actuator $k$ normalized to $d^\mathrm{norm}$, $g_k$ denotes Rayleigh fading with $g_k\sim\mathcal{C}\mathcal{N}(0,1)$, and $\mu$ describes the path loss exponent \cite{9174163}. From \eqref{eq:yk}, we observe that an actuator $k$ receives its own intended signal (first term), the signals from all other sensors $i \neq k$ (second term), and noise (third term). This means that each actuator receives cross-link interference, as can be seen in Fig.~\ref{fig:network}. In this work, we especially apply the treating interference as noise (TIN) scheme, which is shown to be optimal in low-interference scenarios \cite{6824745}. 
	Under the TIN scheme, the instantaneous achievable rate of sensor-actuator pair $k$ is\vspace*{-.25cm}
	\begin{align}\label{eq:rk}
	    r_k = B\,\log_2\left(1+\Gamma_k\right),
	\end{align}
	\\\phantom{a}\vspace*{-1.1cm}\\
	with an SINR of\vspace*{-.2cm}
	\begin{align}\label{eq:sinr}
	    \Gamma_k = \frac{|h_{k,k}|^2 q_k}{\sigma^2 + \sum_{i\neq k} |h_{i,k}|^2 q_i}.
	\end{align}
	\\\phantom{a}\vspace*{-.85cm}\\
	We let $\bm{r}$ be a vector containing all user rates, i.e., $\bm{r}=[r_1,\cdots,r_K]^T$. However, the lack of full CSIT knowledge prohibits the utilization of \eqref{eq:rk} directly. In this contribution, only statistical knowledge of the channel is assumed. That is, adopting $r_k$ without proper channel knowledge produces unforseeable outages, as any value of $r_k$ higher than the instantaneous achievable rate in \eqref{eq:rk} means an outage. The outage probability of sensor-actuator pair $k$ is expressed as\vspace{-.2cm}
	\begin{align}
	    P_k^\mathrm{out}({r}_k) &= P\left( r_k > B\,\log_2(1+\Gamma_{k}) \right)\nonumber\\
	    &= 1 - P\left( r_k \leq B\,\log_2(1+\Gamma_{k}) \right).
	    \label{eq:outageprob}
	\end{align}
	\\\phantom{a}\vspace*{-1cm}\\
	The rationale behind equation \eqref{eq:outageprob} is that the link $k$, i.e., the link between sensor and actuator $k$, is in outage, if the maximum achievable rate falls below the allocated rate $r_k$.\\
	\indent Along with the outage probability of packet transmission, the actual transmission time of the packets plays a major role in the analysis of the AoI in the considered network. Hence, the transmission delay $t_k$ of $k$ can be related as\vspace*{-.2cm}
	\begin{equation}
	    t_k = {N_k}/{r_k},\label{eq:t_k}
	\end{equation}
	\\\phantom{a}\vspace*{-1.1cm}\\
	where $N_k$ is the data size and $r_k$ is the allocated data rate. Note that such relation only captures the transmission delay, i.e., the time loss of the physical sensor-actuator link. If $t_k$ is greater than the coherence time of one channel realization, \eqref{eq:t_k} has to be validated for multiple channel realization. That is, $k$ is in outage if at least one channel realization within $t_k$ renders $r_k$ non-achievable.\\
	\indent After one packet is transmitted successfully, the transmitter continues instantaneously sending update packets. Hence, the peak AoI, i.e., the maximum values of the AoI metric, of a sensor-actuator pair $k$ is denoted as\vspace*{-.25cm}
	\begin{align}
	    \tau_k(l)=t_k+u_k(l),\label{eq:peakAoiComponents}
	\end{align}
	\\\phantom{a}\vspace*{-1.1cm}\\
	in which $u_k(l)$ is the time until the successor of packet $l$ arrives at the actuator. Based on above descriptions, we next proceed to describe the optimization problem in details.\vspace*{-.1cm}
	
	\section{Problem Formulation}\label{sec:prob}\vspace*{-.1cm}
	In this work, we focus on the minimization of the peak AoI among sensor-actuator pairs with different criticality levels, e.g., as considered in \cite{9302619}. 
	That is, for tractability reasons we let such pairs be either of low criticality, i.e., $k\in\mathcal{K}^\text{LO}$ denotes non-critical links, or of high criticality, i.e., $k\in\mathcal{K}^\text{HI}$ denotes safety-critical links. Such association allows the network to prioritize high critical links, if their peak AoI values increase. An objective function acccounting for the peak AoI and the criticality levels is given as\vspace{-.2cm}
    \begin{align}
	    \Psi(\bm{t},\bm{p})=\sum_{k\in\mathcal{K}^\text{LO}}\mathds{E}_l\left[\frac{\tau_k(l)}{\Bar{\tau}}\right]+\sum_{k\in\mathcal{K}^\text{HI}}\mathds{E}_l\left[2^{\tau_k(l)/\Bar{\tau}}\right],\label{eq:objectiveAvPeak1}
	\end{align}
	\\\phantom{a}\vspace*{-.8cm}\\
	where $\Bar{\tau}$ is a normalization factor. The first term in \eqref{eq:objectiveAvPeak1} captures the expected peak AoI of low critical users, i.e., the age is a linear function, whereas the second term in \eqref{eq:objectiveAvPeak1} refers to high critical users with an exponential aging function. Such functions typically depend on $(a)$ the packet outage probability and $(b)$ the packet transmission times, which is dependent on the data size and data rate.\\
	\indent	An optimization problem to minimize the average network-wide peak AoI metric in \eqref{eq:objectiveAvPeak1} under transmission time and outage probability constraints is mathematically given as follows\vspace*{-.2cm}%
	\begin{subequations}\label{eq:Opt1}%
    \begingroup
    \addtolength{\jot}{-.1cm}
		\begin{align}
			\underset{\bm{t},\bm{p}}{\mathrm{min}}\quad &\Psi(\bm{t},\bm{p}) \tag{\ref{eq:Opt1}} \\
			\mathrm{s.t.} \quad & r_k = {N_k}/{t_k},  &\forall k &\in \mathcal{K}, \label{eq:tr_relation}\\
			& P_k^\mathrm{out}({r}_k) \leq p_k,  &\forall k &\in \mathcal{K}. \label{eq:outp}
		\end{align}\endgroup%
	\end{subequations}%
	\\\phantom{a}\vspace*{-1.1cm}\\
	The optimization variables included in this problem are the vectors describing the transmission times $\bm{t}=[t_1,\cdots,t_k]^T$ and the outage probabilities $\boldsymbol{p}=[p_1,\cdots,p_k]^T$. Note that constraint \eqref{eq:tr_relation} relates the transmission times and each user's data rate, whereas $r_k$ is determined by knowing $t_k$. 
	
	By jointly managing transmission time and outage probability, problem \eqref{eq:Opt1} minimizes $\Psi(\bm{t},\bm{p})$ subject to the following constraints: \eqref{eq:tr_relation} the transmission time constraint, which relates the transmission time to the data rate and size; and \eqref{eq:outp} the outage probability constraints, which restricts $P_k^\mathrm{out}({r}_k)$ to be less or equal than $p_k$. Finding a solution to problem \eqref{eq:Opt1} is not straightforward since the objective function $\Psi(\bm{t},\bm{p})$ is in general form and needs to be parametrized. Further, the outage probability constraint \eqref{eq:outp} is in a non-convex and stochastic form, which needs further considerations.
	Henceforth, we proceed to reformulate the problem.\vspace*{-.1cm}
	
	\section{Age of Information Optimization}\vspace*{-.1cm}
	Now, we first provide a closed-form expression of $\Psi(\bm{t},\bm{p})$ and proceed afterwards by formulating a convex optimization problem and an algorithm to solve the original problem.\vspace*{-.1cm}
	\subsection{Average Peak AoI and Mixed-Criticality}\vspace*{-.1cm}
	The objective \eqref{eq:Opt1} contains the expectation over all achieved peak AoI values. To simplify the objective, we reformulate it as presented next. As stated in \eqref{eq:peakAoiComponents}, the peak AoI consists of the transmission time and the waiting time until the follow-up packet is received. The transmission of the follow-up packets of the sensor-actuator pair $k$ fails with probability $p_k$, which is independent for each transmission. As multiple successive transmissions can fail in a row, the peak AoI of the different packets $l$ of pair $k$ is distributed as (see also \cite{9646490})\vspace*{-.2cm}
	\begin{align}
	    \tau_k(l)\sim\sum_{v=0}^{\infty}p_k^v\left(1-p_k\right)\delta\left(\tau_k-(2+v)t_k\right),\label{peakAoIdist}
	\end{align}
	\\\phantom{a}\vspace*{-.8cm}\\
	which is i.i.d. for each packet, and where $\delta(\cdot)$ is the Dirac delta function.
	We can calculate the expectations of the linear and exponential functions of the AoI in \eqref{eq:objectiveAvPeak1} using \eqref{peakAoIdist} as follows. For the linear function of the AoI, we get\vspace*{-.2cm}
    \begingroup
    \addtolength{\jot}{-.08cm}
	\begin{align}
	    \mathds{E}_l\left[\frac{\tau_k(l)}{\Bar{\tau}}\right]&\stackrel{(a)}{=}\frac{1}{\Bar{\tau}}\sum_{v=0}^{\infty}(2+v)t_kp_k^v\left(1-p_k\right)\\
	    &\stackrel{(b)}{=}\frac{t_k\left(1-p_k\right)}{\Bar{\tau}}\left[2\sum_{v=0}^{\infty}p_k^v+\sum_{v=0}^{\infty}vp_k^v\right]\\
	    &\stackrel{(c)}{=}\frac{t_k\left(1-p_k\right)}{\Bar{\tau}}\left[\frac{2}{1-p_k}+\frac{p_k}{\left(1-p_k\right)^2}\right]\\
	    &=2\frac{t_k}{\Bar{\tau}}+\frac{p_k}{1-p_k}\frac{t_k}{\Bar{\tau}}.\label{eq:expectation_linear}
	\end{align}\endgroup
	\\\phantom{a}\vspace*{-.8cm}\\
	In step $(a)$, we utilize the characteristics of the Dirac delta function, in $(b)$, the equation is re-written, and in $(c)$, we utilize the geometric series.
	For the exponential function of the AoI instead, we obtain\vspace*{-.2cm}
    \begingroup
    \addtolength{\jot}{-.1cm}
	\begin{align}
	    \mathds{E}_l\left[2^{\tau_k(l)/\Bar{\tau}}\right]&\stackrel{(a)}{=}\sum_{v=0}^{\infty}2^{(2+v)t_k/\Bar{\tau}}p_k^v\left(1-p_k\right)\\
	    &\stackrel{(b)}{=}2^{2t_k/\Bar{\tau}}\left(1-p_k\right)\sum_{v=0}^{\infty}\left(2^{t_k/\Bar{\tau}}p_k\right)^\nu\\
	    &\stackrel{(c)}{=}\frac{2^{2t_k/\Bar{\tau}}\left(1-p_k\right)}{1-2^{t_k/\Bar{\tau}}p_k},\label{eq:expectation_exponential}
	\end{align}\endgroup
	where, again, in $(a)$, the Dirac delta characteristics are used, $(b)$ is re-writing, and in $(c)$, the geometric series is applied.
	To ensure that the geometric series converges, we introduce
	the constraint $2^{t_k/\tau}p_k<1$. 
	Using the two expectations in \eqref{eq:expectation_linear} and \eqref{eq:expectation_exponential}, we can reformulate the objective \eqref{eq:objectiveAvPeak1} to\vspace*{-.2cm}
    \begingroup
    \addtolength{\jot}{-.4cm}
	\begin{align}
	    \Psi(\bm{t},\bm{p})=\hspace*{-.1cm}\sum_{k\in\mathcal{K}^\text{LO}}\hspace*{-.1cm}\bigg[\underbrace{2\frac{t_k}{\Bar{\tau}}\phantom{\bigg]}\hspace*{-.25cm}}_{(i)}+\underbrace{\frac{p_k}{1-p_k}\frac{t_k}{\Bar{\tau}}\phantom{\bigg]}\hspace*{-.25cm}}_{(ii)}\bigg]+\sum_{k\in\mathcal{K}^\text{HI}}\underbrace{\frac{2^{2t_k/\Bar{\tau}}\left(1-p_k\right)}{1-2^{t_k/\Bar{\tau}}p_k}\phantom{\bigg]}\hspace*{-.25cm}}_{(iii)}.\nonumber\\
	    \label{eq:objectiveAvPeak2}
	\end{align}\endgroup
	\\\phantom{a}\vspace*{-1cm}\\
	Moreover, problem \eqref{eq:Opt1}, with the additional convergence constraint, can be re-written as\vspace*{-.2cm}
	\begin{subequations}\label{eq:Opt2}
    \begingroup
    \addtolength{\jot}{-.09cm}
		\begin{align}
			\underset{\bm{t},{\bm{p}}}{\mathrm{min}}\quad &\Psi(\bm{t},\bm{p}) \tag{\ref{eq:Opt2}} \\
			\mathrm{s.t.} \quad & \eqref{eq:tr_relation}, \eqref{eq:outp}, \nonumber\\
			& 2^{t_k/\Bar{\tau}}p_k<1.\label{eq:geomcond}
		\end{align}\endgroup
	\end{subequations} 
	\\\phantom{a}\vspace*{-1cm}\\
	According to \eqref{eq:outp}, $p_k$ is an upper bound on the outage probability. Problem \eqref{eq:Opt2} has a non-convex objective and non-convex constraints \eqref{eq:outp} and \eqref{eq:geomcond}, which prohibits deriving solutions efficiently. Therefore, we proceed to reformulate problem \eqref{eq:Opt2} utilizing techniques from optimization theory in order to derive a convex optimization problem solvable using low-complexity methods.
	
	\subsection{Convexification of Problem \eqref{eq:Opt2}}\vspace*{-.1cm}
	The reformulation steps to convexify problem \eqref{eq:Opt2} follow a specific order given as follows: $(a)$ The non-convex and fractional nature of the objective function $\Psi(\bm{t},\bm{p})$ in \eqref{eq:objectiveAvPeak2} is relaxed using FP and an SCA approach; $(b)$ The outage probability constraint \eqref{eq:outp} is given in closed-form which is consequently tackled using SCA; $(c)$ The regulation constraint \eqref{eq:geomcond} is reformulated under a similar SCA approach. Each step introduces auxiliary variables which either contribute to the set of optimization variables or are updated in an outer loop.
	
	\subsubsection{FP and SCA for \eqref{eq:objectiveAvPeak2}}
	While the term $(i)$ in \eqref{eq:objectiveAvPeak2} poses no challenge, the terms $(ii)$ and $(iii)$ are of fractional nature.
	Further, solving both terms becomes involved due to the non-linear relation of optimization variables. To first decouple nominator and denominator in those fractional terms, we utilize fractional programming, especially the quadratic transform, as proposed in \cite[Theorem 1]{8314727}. Hence, we obtain\vspace*{-.2cm} 
    \begingroup
    \addtolength{\jot}{-.09cm}
	\begin{align}
	    \Psi(\bm{t},\bm{p})&=\sum_{k\in\mathcal{K}^\text{LO}}\Bigg[\frac{2}{\Bar{\tau}}t_k
	    +\underbrace{{2} \alpha_k \sqrt{p_k \frac{t_k}{\Bar{\tau}}}\phantom{\bigg]}\hspace*{-.25cm}}_{(i)}
	    -\alpha_k^2 
	    +\alpha_k^2 p_k\Bigg] \label{eq:objectiveAvPeak4}\\
	    &\hspace{0.2cm}+ \sum_{k\in\mathcal{K}^\text{HI}}\bigg[ \underbrace{2 \beta_k \sqrt{2^{{2}t_k/{\Bar{\tau}}}(1-p_k)}\phantom{\bigg]}\hspace*{-.25cm}}_{(ii)} 
	    -\beta_k^2 
	    +\underbrace{\beta_k^2 2^{t_k/\Bar{\tau}}p_k\phantom{\bigg]}\hspace*{-.25cm}}_{(iii)}\bigg]
	    ,\nonumber
	\end{align}\endgroup
	\\\phantom{a}\vspace*{-.8cm}\\
	where $\alpha_k$ and $\beta_k$ are auxiliary variables introduced by the quadratic transform. When fixing such auxiliary variables, most terms in \eqref{eq:objectiveAvPeak4} become convex, or need to be further tackled by relaxing the variable coupling between $p_k$ and $t_k$. Keeping all other variables fixed, however, we express the optimal auxiliary variables by \vspace*{-.2cm}
    \begin{align}
         \alpha_k &= \frac{\sqrt{\tilde{p}_k \Tilde{t}_k/{\Bar{\tau}}}}{1-\tilde{p}_k}; & \beta_k &= \frac{\sqrt{2^{2\Tilde{t}_k/{\Bar{\tau}}}(1-\tilde{p}_k)}}{1-2^{\Tilde{t}_k/{\Bar{\tau}}}\tilde{p}_k}\label{eq:alphabeta},
    \end{align}
	where $\tilde{p}_k$ and $\Tilde{t}_k$ are the optimal values obtained within the previous iteration, also referred to as feasible fixed values. Going back to equation \eqref{eq:objectiveAvPeak4}, we note that the terms $(i)$-$(iii)$ 
	remain complicated and a reason for the non-convexity of \eqref{eq:objectiveAvPeak4}. In order of their appearance, we next apply the SCA framework to these terms. More specifically, each term is approximated by its first order Taylor expansion around an operating point, i.e., feasible fixed values. The first term $(i)$, i.e., $\sqrt{p_k t_k/{\Bar{\tau}}}$, can be approximated by \vspace*{-.2cm}
	\begin{align}
	    \sqrt{\tilde{p}_k \tilde{t}_k/{\Bar{\tau}}} + \frac{\Tilde{t_k}/{\Bar{\tau}}(p_k-\tilde{p}_k)}{2\sqrt{\tilde{p}_k \tilde{t}_k/{\Bar{\tau}}}} + \frac{\tilde{p}_k/{\Bar{\tau}}(t_k-\Tilde{t}_k)}{\sqrt{\tilde{p}_k \tilde{t}_k/{\Bar{\tau}}}}.
	\end{align}
	The term $(ii)$, i.e., $\sqrt{2^{{2}t_k/{\Bar{\tau}}}(1-p_k)}$, is approximated by \vspace*{-.2cm}
	\begin{align}
	    2^{\tilde{t}_k/{\Bar{\tau}}} \Bigg(\sqrt{(1-\tilde{p}_k)} \Big( 1 + \frac{\text{ln}(2)}{\Bar{\tau}}(t_k-\tilde{t_k})\Big)
	    -\frac{p_k-\tilde{p}_k}{2\sqrt{(1-\tilde{p}_k)}}\Bigg).
	\end{align}
	\\\phantom{a}\vspace*{-.8cm}\\
	The last term $(iii)$ in \eqref{eq:objectiveAvPeak4}, i.e., $2^{t_k/\Bar{\tau}}p_k$, becomes\vspace*{-.2cm}
	\begin{align}
	    2^{\tilde{t}_k/\Bar{\tau}}\tilde{p}_k + \frac{\text{ln}(2)}{\Bar{\tau}}\tilde{p}_k 2^{\tilde{t}_k/\Bar{\tau}}(t_k-\tilde{t}_k) + 2^{\tilde{t}_k/\Bar{\tau}} (p_k-\tilde{p}_k).
	\end{align}
	\\\phantom{a}\vspace*{-1cm}\\
	Given the above consideration, the reformulated and approximated objective function $\Psi(\bm{t},\bm{p})$ may now be written into \eqref{eq:objectiveAvPeak5}, as can be seen on top of the next page in details.
	\begin{figure*}[t]
	\normalsize
	\begin{align}
	    \Psi&(\bm{t},\bm{p})=\sum_{k\in\mathcal{K}^\text{LO}}\Bigg[\frac{2}{\Bar{\tau}}t_k
	    +2 \alpha_k \left[\sqrt{\tilde{p}_k \tilde{t}_k/{\Bar{\tau}}} + \frac{\Tilde{t_k}/{\Bar{\tau}}(p_k-\tilde{p}_k)}{2\sqrt{\tilde{p}_k \tilde{t}_k/{\Bar{\tau}}}} + \frac{\tilde{p}_k/{\Bar{\tau}}(t_k-\Tilde{t}_k)}{\sqrt{\tilde{p}_k \tilde{t}_k/{\Bar{\tau}}}}\right]
	    -\alpha_k^2 
	    +\alpha_k^2 p_k\Bigg] + \sum_{k\in\mathcal{K}^\text{HI}} \Bigg[2 \beta_k \nonumber\\
	    &\times\left[2^{\tilde{t}_k/{\Bar{\tau}}} \Bigg(\sqrt{(1-\tilde{p}_k)} \Big( 1 + \frac{\text{ln}(2)}{\Bar{\tau}}(t_k-\tilde{t_k})\Big)
	    -\frac{p_k-\tilde{p}_k}{2\sqrt{(1-\tilde{p}_k)}}\Bigg)\right]
	    -\beta_k^2 
	    +\beta_k^2 \left[ \frac{\text{ln}(2)}{\Bar{\tau}}\tilde{p}_k 2^{\tilde{t}_k/\Bar{\tau}}(t_k-\tilde{t}_k) + 2^{\tilde{t}_k/\Bar{\tau}} p_k\right]\Bigg]
	    \label{eq:objectiveAvPeak5}
	\end{align}
	\hrulefill
	\vspace*{-.6cm}
\end{figure*}%
    \subsubsection{SCA for Outage Probability Constraint \eqref{eq:outp}}\label{ssec:outprobSCA}
    Based on \cite{6365845} and \cite{975444}, the outage probability of D2D pair $k$ can be expressed as the following closed-form equation\vspace*{-.2cm}
	\begin{align}
	    P_k^\mathrm{out}({r}_k) &= \nonumber\\
	    &\hspace{-0.95cm}1 -  e^{-\frac{\left(2^{r_k/B}-1\right)d_{k,k}^{\mu/2}\sigma_k^2}{q_k}} \prod_{i\neq k} \frac{q_k}{q_k+\left(2^{r_k/B}-1\right)\frac{d_{k,k}^{\mu/2}}{d_{i,k}^{\mu/2}}q_i} .\label{eq:outageprob_}
	\end{align}
	\\\phantom{a}\vspace*{-.85cm}\\
    The rationale behind such formulation is the intricate coupling between D2D pairs, i.e., the interference among each link, and additionally the noise. All these factors are relevant to the outage probability of user $k$ (for more details and proofs, please refer to \cite{975444}).
    Using some mathematical manipulations, 
    the outage probability constraint \eqref{eq:outp} (using \eqref{eq:outageprob_}) becomes\vspace*{-.2cm}
	\begin{equation}
	    (1-p_k) e^{(2^{r_k/B}-1)\sigma_k^2\frac{d_{k,k}^{\mu/2}}{q_k}} \prod_{i\neq k} \left(1+(2^{r_k/B}-1)\frac{q_i d_{k,k}^{\mu/2}}{q_k d_{i,k}^{\mu/2}}\right) \leq 1. \label{eq:outageprob2}
	\end{equation}
	\\\phantom{a}\vspace*{-.9cm}\\
    In order to tackle the non-convex outage probability constraint \eqref{eq:outageprob2}, we firstly define the following auxiliary variables\vspace*{-.2cm}
    \begin{align}
        e^{y_k} &= 2^{r_k/B} - 1; & e^{x_{i,k}} &= \frac{q_i}{d_{i,k}^{\mu/2}}; & z_k &= \frac{(2^{r_k/B} - 1)d_{k,k}^{\mu/2}}{q_k}.\nonumber
    \end{align}
	\\\phantom{a}\vspace*{-.9cm}\\
	Note that $z_k$ can also be expressed as $z_k = e^{y_k-x_{k,k}}$, and that $e^{x_{i,k}}$ is fixed as it does not depend on any optimization variable, but is still useful for the following reformulations. Introducing such variables, we may now express the outage probability constraint \eqref{eq:outageprob2} as\vspace*{-.2cm}
	\begin{equation}
	    (1-p_k) e^{\sigma_k^2 z_k} \prod_{i\neq k} \Big(1+e^{-x_{k,k}+x_{i,k}+y_k}\Big) \leq 1. \label{eq:outageprob3}
	\end{equation}
	\\\phantom{a}\vspace*{-.8cm}\\
	However, \eqref{eq:outageprob3} is still challenging to solve, thus, we introduce another auxiliary variable $a_k$, replacing the complex exponential and multiplication terms. Let $\bm{a} = [a_1,\cdots,a_K]^T$, $\bm{z} = [z_1,\cdots,z_K]^T$, and $\bm{y} = [y_1,\cdots,y_K]^T$ be part of the set of optimization variables, which additionally introduce the following constraints, for all $k\in\mathcal{K}$:\vspace*{-.3cm}
    \begingroup
    \addtolength{\jot}{-.1cm}
	\begin{align}
	    a_k - p_k a_k -1 &\leq 0, \label{eq:outageprob4}\\
	    r_k - B \log_2(1+e^{y_k}) &\leq 0, \label{eq:rte_yk}\\
	    e^{y_k-x_{k,k}} - z_k &\leq 0, \label{eq:ykzk}\\
	    e^{\sigma_k^2 z_k} \prod_{i\neq k} \Big(1+e^{-x_{k,k}+x_{i,k}+y_k}\Big) - a_k &\leq 0.\label{eq:a_k}
	\end{align}\endgroup
	\\\phantom{a}\vspace*{-.85cm}\\
	Note that constraints \eqref{eq:outageprob4} and \eqref{eq:rte_yk} need further considerations to be utilized in a convex solver. Therefore, we utilize an SCA approach to reformulate \eqref{eq:rte_yk} into\vspace*{-.2cm} 
	\begin{align}
	    r_k - B \log_2(1+e^{\tilde{y}_k}) - \frac{B e^{\tilde{y}_k}}{\text{ln}(2)(1+e^{\tilde{y}_k})}(y_k-\tilde{y}_k) &\leq 0. \label{eq:rte_yk2}
	\end{align}
	\\\phantom{a}\vspace*{-.85cm}\\
	To tackle the constraint \eqref{eq:outageprob4}, which is of bilinear nature, we equally reformulate $-p_k a_k$ as $-\frac{1}{4}((p_k+a_k)^2-(p_k-a_k)^2)$. Approximating the non-convex part, i.e., $-\frac{1}{4}(p_k+a_k)^2$, using the first order Taylor expansion, we obtain\vspace*{-.2cm}
    \begingroup
    \addtolength{\jot}{-.1cm}
	\begin{align}
	    a_k + &\frac{1}{4}(p_k-a_k)^2 - \frac{1}{4}(\tilde{p}_k+\tilde{a}_k)^2 \nonumber\\
	    &- \frac{1}{2}(\tilde{p}_k+\tilde{a}_k)^2 \big[(p_k-\tilde{p}_k)+(a_k-\tilde{a}_k)\big] - 1 \leq 0, \label{eq:outageprob5}
	\end{align}\endgroup
	\\\phantom{a}\vspace*{-.95cm}\\
	which is of convex form. The above reformulations introduce the additional optimization variables $\bm{a}$, $\bm{z}$, and $\bm{y}$. Also, $\bm{\tilde{y}}=[\tilde{y}_1,\cdots,\tilde{y}_K]^T$, 
	$\bm{\tilde{t}}=[\tilde{t}_1,\cdots,\tilde{t}_K]^T$, $\boldsymbol{\tilde{p}}=[\tilde{\epsilon}_1,\cdots,\tilde{p}_k]^T$, and  $\bm{\tilde{a}}=[\tilde{a}_1,\cdots,\tilde{a}_K]^T$ denote feasible fixed values updated in an outer loop.
	
	\subsubsection{SCA for Regulation Constraint \eqref{eq:geomcond}}
	As a final reformulation step, consider the regulation constraint \eqref{eq:geomcond}, which can be approximated using the SCA framework by\vspace*{-.2cm}
	\begin{align}
	    2^{\tilde{t}_k/\Bar{\tau}}\tilde{p}_k+\frac{\text{ln}(2)}{\Bar{\tau}} 2^{\tilde{t}_k/\Bar{\tau}}\tilde{p}_k (t_k-\tilde{t}_k)+2^{\tilde{t}_k/\Bar{\tau}} (p_k-\tilde{p}_k)<1.\label{eq:geomcond2}
	\end{align}
	\\\phantom{a}\vspace*{-.85cm}\\
	Hence, at this point, problem \eqref{eq:Opt2}'s non-convexity is tackled using FP and SCA techniques, making the problem amenable for efficient solution approaches.\vspace*{-.1cm}
	
	\subsection{Convex Problem and Algorithm}\vspace*{-.1cm}
	Given all above considerations and reformulations, the convex relaxed optimization problem can be mathematically formulated as\vspace*{-.2cm}
	\begin{subequations}\label{eq:Opt3}
    \begingroup
    \addtolength{\jot}{-.1cm}
		\begin{align}
			\underset{\bm{t},{\bm{p}},\bm{y},\bm{z},\bm{a}}{\mathrm{min}}\quad &\Psi(\bm{t},\bm{p}) \tag{\ref{eq:Opt3}} \\
			\mathrm{s.t.} \;\;\;\quad & \eqref{eq:tr_relation}, \eqref{eq:ykzk}, \eqref{eq:a_k}, \eqref{eq:rte_yk2}, \eqref{eq:outageprob5}, \eqref{eq:geomcond2}.\nonumber
		\end{align}\endgroup
	\end{subequations} 
	\\\phantom{a}\vspace*{-1cm}\\
	\begin{algorithm}[b]
    \caption{Minimizing Mixed-Critical $\Psi(\bm{t},\bm{p})$}
    \begin{algorithmic}[1]
	\STATE Initialize feasible fixed values $\bm{\tilde{y}}$, $\boldsymbol{\tilde{p}}$, $\bm{\tilde{a}}$, and $\bm{\tilde{t}}$\vspace*{-.05cm}
	\REPEAT
	\STATE Compute $\boldsymbol{\alpha}$ and $\boldsymbol{\beta}$ using \eqref{eq:alphabeta}\vspace*{-.05cm}
	\STATE Solve the convex problem \eqref{eq:Opt3}\vspace*{-.05cm}
	\STATE Update $\bm{\tilde{y}}$, $\boldsymbol{\tilde{p}}$, $\bm{\tilde{a}}$, and $\bm{\tilde{t}}$ \vspace*{-.05cm}
	\UNTIL{convergence}
    \end{algorithmic}
    \label{alg}
    \end{algorithm}%
	As problem \eqref{eq:Opt3} has a convex objective function and convex constraints, \eqref{eq:Opt3} is a convex problem which can be solved using CVX \cite{cvx}. Due to the variety of feasible fixed values and quadratic transform variables, which are updated iteratively, Algorithm~\ref{alg} provides detailed steps for the resource management to minimize the mixed-critical objective function $\Psi(\bm{t},\bm{p})$ in the considered system model. More specifically, initially, Algorithm~\ref{alg} initializes the feasible fixed values $\bm{\tilde{y}}$, $\boldsymbol{\tilde{p}}$, $\bm{\tilde{a}}$, and $\bm{\tilde{t}}$. As this is not straightforward, we provide our approach to initialize these values next. Based on the statistical CSIT, we may generate $M$ channel realizations and compute the rate vector $\bm{r}$ using \eqref{eq:rk} of the worst-case channel realization. If $M$ is a sufficiently large integer, the worst-case approximation of $r_k$ is feasible for almost all cases. From the worst-case $\bm{r}$, we may calculate $\bm{\tilde{t}}$ as \eqref{eq:t_k}, $\boldsymbol{\tilde{p}}$ by setting the inequality as an equality in \eqref{eq:outp}, $\bm{\tilde{y}}$ according to Section~\ref{ssec:outprobSCA}, and $\bm{\tilde{a}}$ by setting \eqref{eq:a_k} to equality.
	The following steps are repeated until convergence. First, compute the quadratic transform constants $\boldsymbol{\alpha} = [\alpha_1,\cdots,\alpha_K]^T$ and $\boldsymbol{\beta} = [\beta_1,\cdots,\beta_K]^T$. Next, solve the convex optimization problem \eqref{eq:Opt3} using CVX. Lastly, the feasible fixed values are updated according to problem \eqref{eq:Opt3}'s solution.
    \subsubsection{Computational Complexity of Algorithm~\ref{alg}}
    Both the convergence rate and the overall complexity of the optimization problem \eqref{eq:Opt3} are relevant to Algorithm~\ref{alg}'s computational complexity. Problem \eqref{eq:Opt3} is solvable using an interior-point method, with $\xi = 6K$ being the total number of variables \cite{Lobo1998ApplicationsOS}. Hence, the overall upper-bound computational complexity of Algorithm~\ref{alg} is given as $\mathcal{O}({V}_{\text{max}}\xi^{3.5})$, where ${V}_{\text{max}}$ is the worst-case number of iterations until Algorithm~\ref{alg} converges. The performance of Algorithm~\ref{alg} is validated numerically in the following section.
	
\section{Numerical Simulation}\label{sec:sim}
    This section is devoted to demonstrating simulation results for the proposed mixed-critical AoI optimization as per Algorithm~\ref{alg}. We consider an area of $100 \times \SI{100}{m}$. Similar to \cite{6824745}, $K$ D2D transmitter-receiver pairs are placed randomly within the simulation area with a minimum distance of $5$m and a maximum distance of $25$m. Further, the receiver is assumed to be placed at a minimum distance of $20$m from \emph{any other} interfering transmitter. We set the noise power spectral density to $\SI{-134}{dBm/Hz}$, the bandwidth to $B=\SI{10}{MHz}$, the transmit power to $q_k = \SI{20}{dBm}$, $\forall k\in\mathcal{K}$, the pathloss exponent to $\mu=2$, and the normalized distance to $d^\text{norm}=\SI{1}{m}$. Unless mentioned otherwise, let the data size be $N_k=\SI{50}{kbit}$, for each user, and the normalization parameter be $\Bar{\tau}=\SI{10}{s}$.
    
    This paper proposes a NOMA-based scheme, where devices are transmitting simultaneously. Contrary to this, we consider an orthogonal multiple access (OMA) scheme, which can for example assign users to different frequency bands or time slots as a baseline.
    For simplicity, such scheme utilizes Algorithm~\ref{alg} with a small modification. That is, under OMA we share the resources equally among devices, i.e., $B^\text{OMA} = B/K$.
    
    To obtain some further understanding of the system, we first consider the system dynamics over time.
    
    \subsection{Instantaneous AoI over Time}\vspace*{-.1cm}
    \begin{figure*}[t!]
     \centering
     \subfloat[][AoI of the proposed NOMA scheme.]{\includegraphics[width=.33\linewidth]{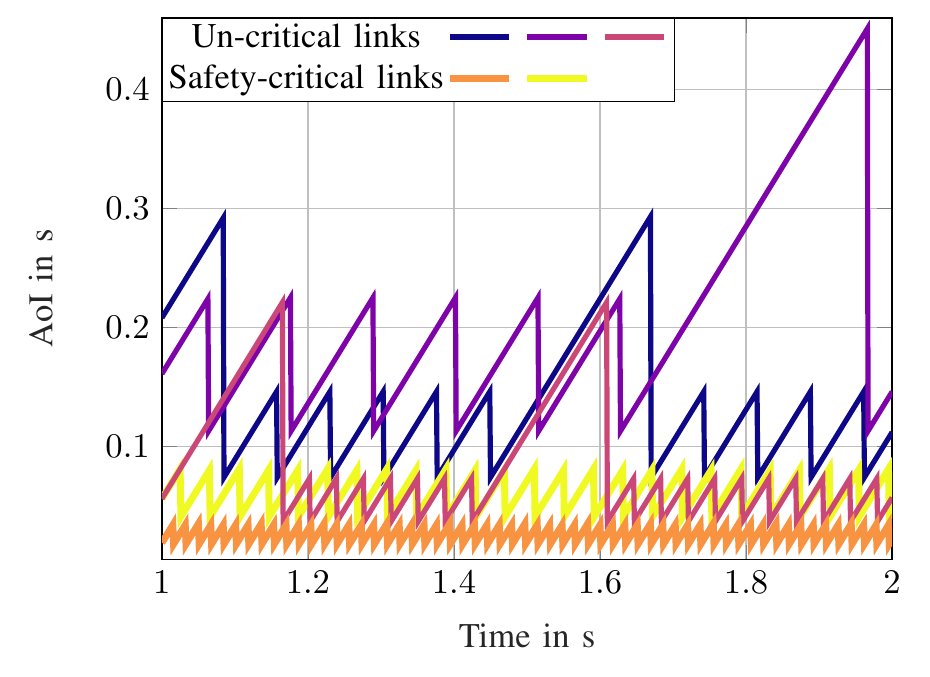}\label{res:F1_yAoI_xTime_NOMA_v2}}
     \subfloat[][AoI of the OMA scheme.]{\includegraphics[width=.33\linewidth]{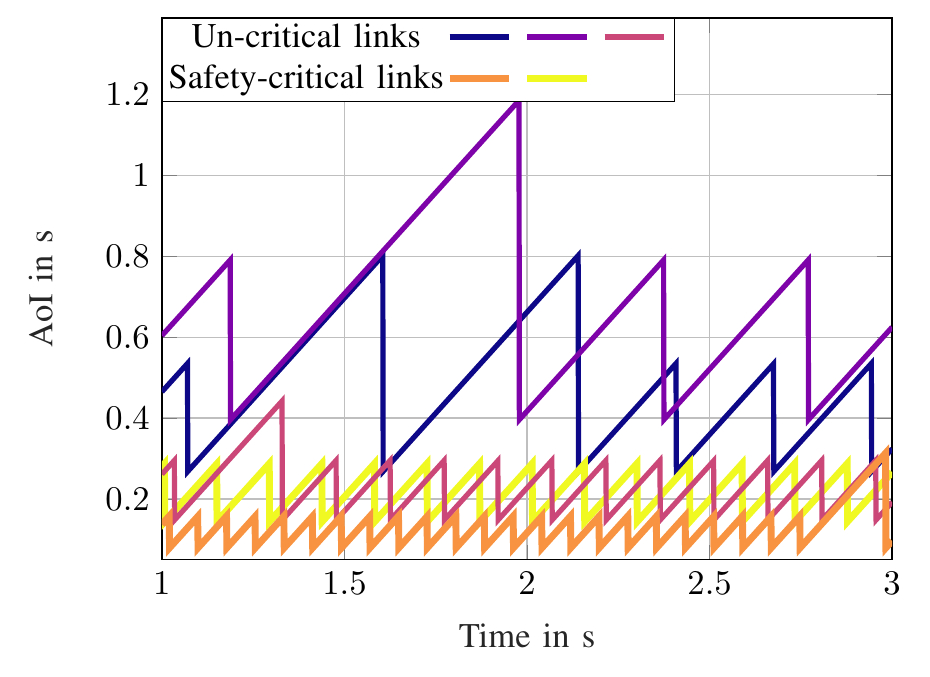}\label{res:F2_yAoI_xTime_OMA_v2}}
     \subfloat[][Optimal and simulated $\Psi(\bm{t},\bm{p})$.]{\includegraphics[width=.33\linewidth]{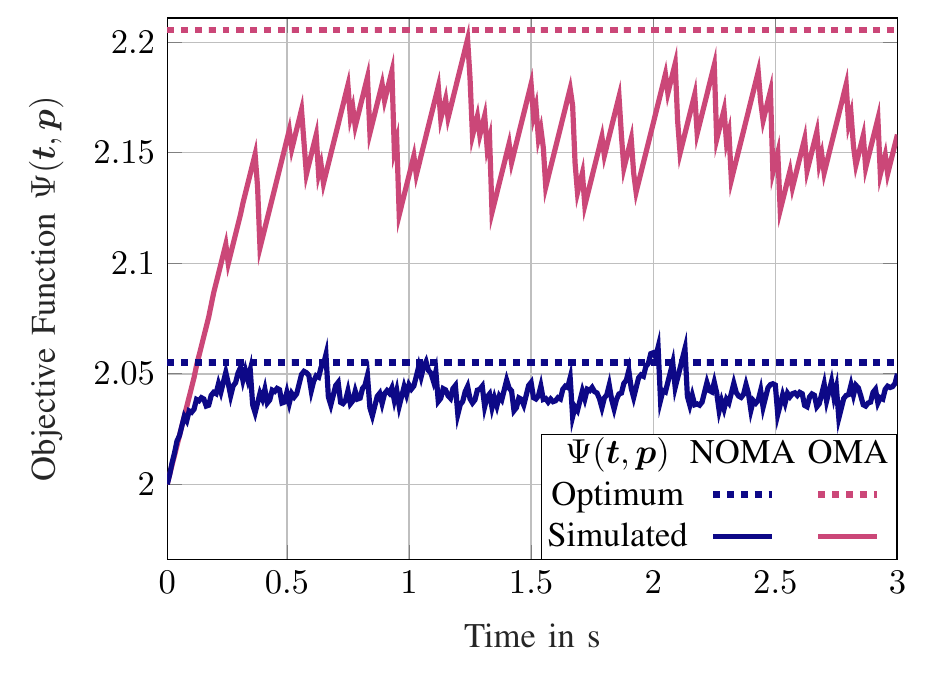}\label{res:F3_yObj_xTime_Compare}}
     \vspace{-.2cm}
     \caption{Instantaneous AoI of the proposed (a) and the OMA (b) scheme, and objective function over time (c).}
     \vspace*{-.7cm}
    \end{figure*}
    We first observe the instantaneous AoI over an interval of $\SI{1}{s}$, where the coherence time is $t^\text{coh}=\SI{150}{ms}$, i.e., the channel realizations change every $\SI{150}{ms}$, while the channel statistics remain constant. We investigate a system with $5$ sensor-actuator pairs and show the AoI of the data available at the actuators in \figurename~\ref{res:F1_yAoI_xTime_NOMA_v2}. The results show the temporal behavior of the different AoI values, which have a significant difference for each actuator. 
    Two out of the five actuators are classified as safety-critical devices. It can be observed from \figurename~\ref{res:F1_yAoI_xTime_NOMA_v2} that these safety-critical devices achieve lower instantaneous AoIs as compared to the non-critical devices. Also, such high priority devices are observed to achieve lower outage probability. 
    In particular, no safety-critical actuator experiences outages in \figurename~\ref{res:F1_yAoI_xTime_NOMA_v2}. Opposed to that, we note that the non-critical links are allocated higher outage probabilities, which results in multiple, even subsequent, packet outages.
    For comparison, \figurename~\ref{res:F2_yAoI_xTime_OMA_v2} plots the same actuators' AoI over an interval of $\SI{2}{s}$ for the OMA transmission scheme. 
    \figurename~\ref{res:F2_yAoI_xTime_OMA_v2} shows that the AoIs approach much higher values as compared to the proposed NOMA transmission scheme. While safety-critical actuators achieve a peak AoI of roughtly $\SI{0.2}{s}$ and $\SI{0.3}{s}$, non-critical AoIs lay beyond these times. In \figurename~\ref{res:F2_yAoI_xTime_OMA_v2}, there is even an actuator which experiences an instantaneous AoI of $\SI{1.2}{s}$. 
    
    Based on these results, the NOMA and OMA schemes' sum mixed-critical peak linear and exponential instantaneous AoI 
    values are shown over time in \figurename~\ref{res:F3_yObj_xTime_Compare} together with the optimized objective function $\Psi(\bm{t},\bm{p})$. 
    Consistent with previous observations, NOMA achieves lower values for both the simulated and optimized $\Psi(\bm{t},\bm{p})$. The curves showing the simulated objective (solid lines) capture the time-varying behavior of the algorithm including simulated channels, coherence time, and outages. In contrast, the optimum curves (dotted lines) show the optimal value according to Algorithm~\ref{alg}. \figurename~\ref{res:F3_yObj_xTime_Compare} shows that the optimum value indeed constitutes a close approximation of the mixed-critical sum of average functions of the AoI. Also, in \figurename~\ref{res:F3_yObj_xTime_Compare}, it becomes clear that the proposed scheme outperforms the OMA benchmark in both optimized values and simulated time-dependent behavior.
    
    To briefly summarize, the above results illustrate the behavior of instantaneous AoI over time, the enhanced performance of the proposed NOMA scheme, and the prioritization of safety-critical over non-critical links.
    %

    \subsection{Impact of Transmit Power and Link Number}
\begin{figure}[t]
		\centering
		\includegraphics[width=.9\linewidth]{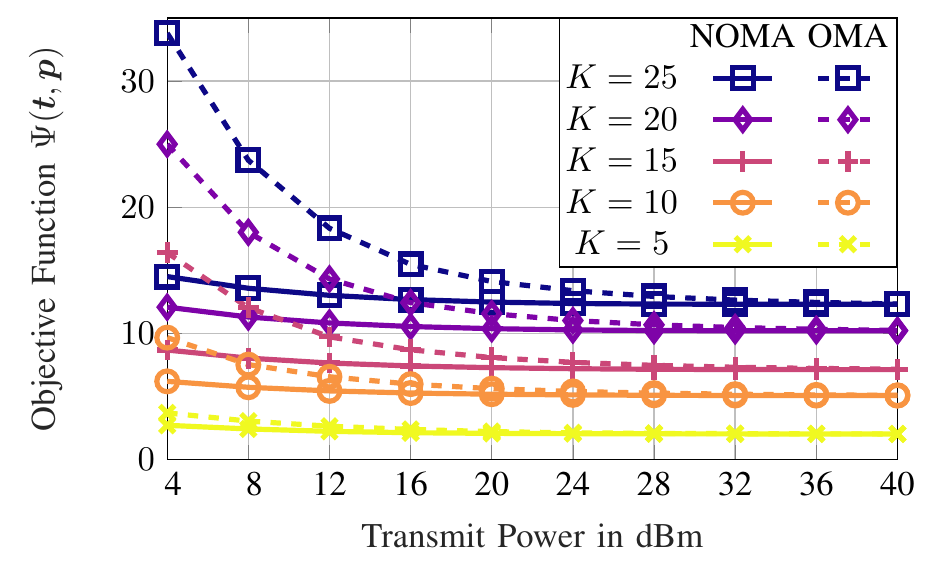}
		\vspace*{-.35cm}
		\caption{Objective function $\Psi(\bm{t},\bm{p})$ over the transmit power, comparing different numbers of links.}
		\label{res:F4_yObj_xTxPow}
	\end{figure}

\vspace*{-.1cm}
    Next, we investigate the impact of various parameters on the objective value. In \figurename~\ref{res:F4_yObj_xTxPow}, the objective function $\Psi(\bm{t},\bm{p})$ for different numbers of links $K$ is shown over the transmit power $q_k$ in dBm.
	First, in \figurename~\ref{res:F4_yObj_xTxPow}, we observe that $\Psi(\bm{t},\bm{p})$ decreases, i.e., improves, for both schemes as the transmit power increases. This decrease is most prominent in the low-power region, since at high power levels the interference becomes the major limiting factor. In particular, \figurename~\ref{res:F4_yObj_xTxPow} shows a saturation of objective values in the high-power regime.
	Interestingly, the number of links plays an equally important role for the objective function. That is, when increasing the number of links, $\Psi(\bm{t},\bm{p})$ increases as well. Such observation is reasonable due to the increased interference level for the NOMA scheme, and due to the reduction of avaliable resources to each device for the OMA scheme.
	Hence, another observation from \figurename~\ref{res:F4_yObj_xTxPow} is the improved scalability  of the proposed NOMA scheme in transmit power limited scenarios. More specifically, when increasing the links, the NOMA scheme's $\Psi(\bm{t},\bm{p})$ increases only slightly, as compared to the OMA scheme. 
	Finally, in \figurename~\ref{res:F4_yObj_xTxPow}, NOMA outperforms OMA up to the interference-heavy scenario, where the transmit power is high. However, in such case, NOMA is still able to perform similarly to OMA, which pronounces the benefits of the proposed algorithm.
	
	Both the enhanced performance and the scalability of the proposed algorithm illustrate its significance to future IoT networks, especially with low transmit powers and large number of links.\vspace*{-.2cm}
	
    
\section{Conclusion}\label{sec:con}\vspace*{-.1cm}
    Use-cases such as industrial control systems and smart transportation pose strict constraints for the underlying communication infrastructure, particularly in terms of swiftly and timely delivering data packets of different criticality. Therefore, this paper considers the freshness of data in mixed-critical D2D sensor-actuator networks, where links have different priorities. Under a NOMA scheme, an optimization problem jointly managing physical transmission time and packet outage probability for minimizing different functions of the AoI is derived. The non-convexity is tackled using stochastic reformulations, SCA, and FP, which results in an efficient iterative procedure for optimizing the AoI. The behavior of instantaneous AoI over time and the impacts of network parameters on the objective are then illustrated in the simulation results. The prioritization of safety-critical data, the improved AoI performance, and the scalability were highlighted, as compared to an orthogonal system strategy. Especially in low transmit power scenarios with high numbers of links, the proposed algorithm demonstrates significant performance gains, which emphasize its suitability for future IoT network applications.\vspace*{-.2cm}
\bibliographystyle{IEEEtran}
\bibliography{bibliography}
\end{document}